\documentclass[fleqn,usenatbib]{mnras}

\usepackage{newtxtext,newtxmath}
\usepackage[T1]{fontenc}
\usepackage{ae,aecompl}
\usepackage{graphicx}
\usepackage{subfigure}

\newcommand{\sag}{\textsc{sag}}

\usepackage{graphicx}	% Including figure files
\usepackage{amsmath}	% Advanced maths commands
\usepackage{epsfig}

\usepackage[dvipsnames]{xcolor}
\title[Associations of dwarf galaxies]%
{Associations of dwarf galaxies in a $\Lambda$CDM Universe}
\author[Yaryura et al.]
{
C.Yamila~Yaryura$^{1}$\thanks{E-mail: yamila.yaryura@unc.edu.ar}, Mario G.~Abadi$^{1,2}$, Stefan~Gottl\"ober$^{3}$,
Noam I.~Libeskind$^{3,4}$, 
\newauthor  Sof\'ia A.~Cora$^{5,6}$, Andr\'es N.~Ruiz$^{1,2}$, Cristian A.~Vega-Mart\'{i}nez$^{7,8}$, 
Gustavo~Yepes $^{9,10}$,
\newauthor Peter~Behroozi$^{11}$
\\
\\
$^{1}$CONICET-Universidad Nacional de C\'{o}rdoba, Instituto de Astronom\'{i}a Te\'{o}rica y Experimental (IATE),
Laprida 854, X5000BGR,C\'{o}rdoba, Argentina.\\
$^{2}$Observatorio Astron\'{o}mico, Universidad Nacional de C\'{o}rdoba, Laprida 854, X5000BGR, C\'{o}rdoba, Argentina.\\
$^{3}$Leibniz-Institut f\"ur Astrophysik Potsdam (AIP), An der Sternwarte 16, D - 14482, Potsdam, Germany. \\
$^{4}$University of Lyon; UCB Lyon 1/CNRS/IN2P3; IPN Lyon, France.\\
$^{5}$Instituto de Astrof\'isica de La Plata (CCT La Plata, CONICET,UNLP), Observatorio Astron\'omico, Paseo del Bosque,\\ B1900FWA, La Plata, Argentina.\\
$^{6}$Facultad de Ciencias Astron\'omicas y Geof\'{\i}sicas, Universidad Nacional de La Plata (UNLP), Observatorio Astron\'omico,\\ Paseo del Bosque, B1900FWA La Plata, Argentina.\\
$^{7}$Instituto de Investigaci\'on Multidisciplinar en Ciencia y Tecnolog\'{\i}a, Universidad de La Serena, Ra\'ul Bitr\'an 1305, La Serena, Chile.\\
$^{8}$Departamento de Astronom\'{\i}a, Universidad de La Serena, Av. Juan Cisternas 1200 Norte, La Serena, Chile.\\
$^{9}$Departamento de F\'{\i}sica Te\'orica M-8, Universidad Aut\'onoma de Madrid, Cantoblanco, E-28049 Madrid, Spain.\\
$^{10}$Centro de Investigaci\'on Avanzada en F\'{\i}sica  Fundamental (CIAFF), Universidad Aut\'onoma de Madrid, E-28049 Madrid, Spain.\\
$^{11}$Department of Astronomy and Steward Observatory, University of Arizona, Tucson, AZ 85721, USA. \\
}

\date{Accepted XXX. Received YYY; in original form ZZZ}

\pubyear{2019}

\begin{document}
\label{firstpage}
\pagerange{\pageref{firstpage}--\pageref{lastpage}}
\maketitle

% Abstract of the paper
\begin{abstract}
Associations of dwarf galaxies are loose systems composed 
exclusively of dwarf galaxies.
These systems were identified in the Local Volume for the first time more than thirty years ago.
We study these systems in the cosmological
framework of the $\Lambda$ Cold Dark Matter ($\Lambda$CDM) model.
We consider the Small MultiDark Planck simulation and populate its dark matter 
haloes by applying the semi-analytic model of galaxy formation SAG.
We identify galaxy systems using a friends of friends algorithm
with a linking length equal to $b=0.4 \,{\rm Mpc}\,h^{-1}$, 
to reproduce the size of dwarf galaxy associations detected in the 
Local Volume.
Our samples of dwarf systems are built up removing those systems 
that have one (or more) galaxies with stellar mass larger than a maximum 
threshold $M_{\rm max}$.
We analyse three different samples defined by 
${\rm log}_{10}(M_{\rm max}[{\rm  M}_{\odot}\,h^{-1}]) = 8.5, 9.0$ and $9.5$. 
On average, our systems have typical sizes of $\sim 0.2\,{\rm Mpc}\,h^{-1}$, 
velocity dispersion of $\sim 30 {\rm km\,s^{-1}} $ and estimated total 
mass of $\sim 10^{11} {\rm  M}_{\odot}\,h^{-1}$.
Such large typical sizes suggest that individual members of a given dwarf
association reside in different dark matter haloes and are generally not 
substructures of any other halo. 
Indeed, in more than 90 per cent of our dwarf systems their
individual members inhabit different dark matter haloes, while only in the remaining 
10 per cent members do reside in the same halo. 
Our results indicate that the $\Lambda$CDM model can naturally reproduce the 
existence and properties of dwarf galaxies associations without much difficulty.

\end{abstract}


%%%%%%%%%%%%%%%%%%%%%%%%%%%%%%%%%%%%%%%%%%%%%%%%%%

%%%%%%%%%%%%%%%%% BODY OF PAPER %%%%%%%%%%%%%%%%%%

\section{Introduction}

In the last two decades, the development of large galaxy surveys such as 
2dF \citep{Colless:2001}, HIPASS \citep{Barnes:2001}, 6dF \citep{Jones:2004},
ALFALFA \citep{Giovanelli:2005}, SDSS \citep{Abazajian:2009}, among others, has
helped clarify our understanding of the large-scale structure of the Universe.
However, the most relevant constraint of these surveys is that they do not 
include numerous dwarf galaxies because of their low luminosity and surface brightness.
In order to deal with this problem, many projects are now focused on the study of 
the nearby Universe within a radius of 10 Mpc, often termed the Local Volume (LV).
The study of the LV allows an accurate estimation of the line--of--sight 
velocities and individual distances for many systems, most of them, dwarf galaxies, 
and enables us to investigate the properties of this galaxy population 
and the interactions between its members.

In the literature, we find numerous studies on the frequency of interactions 
between massive galaxies, 
both from observational and theoretical points of view.
On the contrary, there are not so many studies on the frequency of interactions 
between dwarf galaxies and their role in the evolution of low--mass galaxies.
Although low--mass dwarf galaxies, with 
\mbox{$M_{\rm stellar} < 5 \times 10^9 {\rm M}_{\odot}$}, are the most abundant 
class of galaxies at all redshifts \citep{Binggeli:1988, Karachentsev:2013}, 
there are neither so many 
observational studies about the associations of dwarf galaxies, nor works that compare their 
observed properties with theoretical predictions in a systematic way.

\begin{figure}
  \centering
  \includegraphics[width=0.47\textwidth\vspace{0pt}]{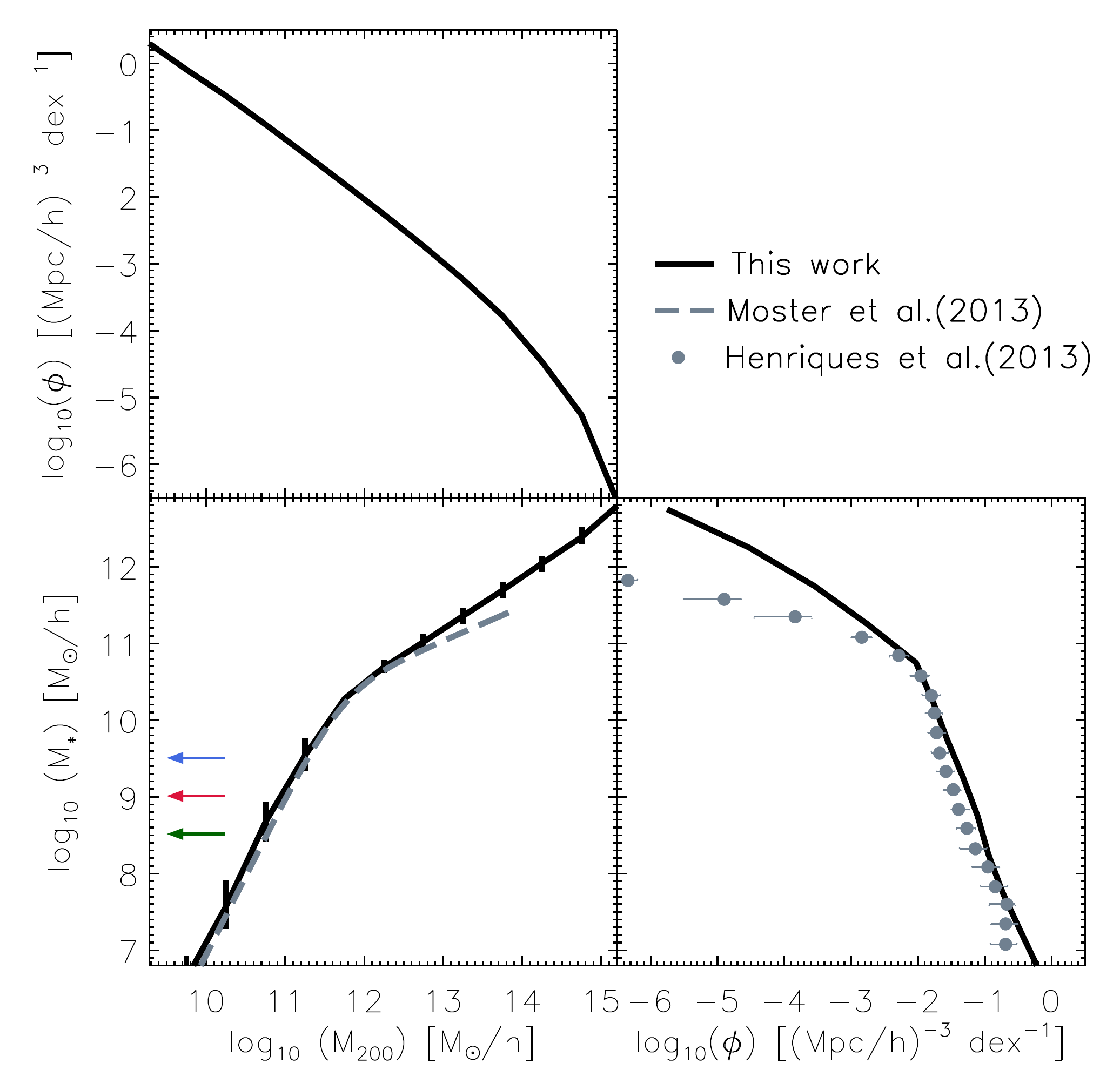}
  \caption{Bottom left panel: Median stellar--virial mass relation 
for all semi--analytical galaxies in the full volume of the SMDPL simulation 
  (black solid line), compared with the correlation obtained from the
  abundance matching technique \citep[][grey dashed line]{Moster:2013}. 
  Present--day halo masses are considered for central galaxies, while 
  infall halo masses are taken into account for satellites. 
  Vertical error bars correspond to $25$ and $75$ per cent quartiles.
  Bottom right panel: Stellar mass function for all
  semi--analytical galaxies in the volume of the SMDPL simulation 
  (black solid line), compared with observational data 
  collected by \citep[][grey filled circles]{Henriques:2013}.
  Top left panel: Virial mass function for all 
  semi--analytical galaxies in the volume of the SMDPL simulation 
  (black solid line).
  Coloured arrows indicate different stellar mass thresholds, 
  $M_{\rm max}$, used to define three different samples of dwarf 
  galaxies systems in sub--section \ref{sample}.} 
\label{fig:SAG}
\end{figure}

We can mention the identification of dwarf galaxy associations presented by 
\cite{Tully:1988} as one of the first studies of dwarf galaxies interactions.
In their work, they define two kinds of structures, called ``groups'' and 
``associations'', according to luminosity density thresholds
that characterize the connection between galaxies.
These thresholds are defined as the ratio between the luminosity of the 
brighter component of a galaxy pair selected by a hierarchical method, 
and the third power of the distance between these two galaxies 
(see \citealt{Tully:1988} for a exhaustive depiction of the method).
They set $\rho_{\rm g} = 2.5 \times 10^9 \, L_{\odot} \,{\rm Mpc}^{-3}$ 
and $\rho_{\rm a} = 2.5 \times 10^8\, L_{\odot}\, {\rm Mpc}^{-3}$ as the density 
thresholds to define a group and an association, respectively. 
Note that the latter is one order of magnitude lower than the former.
They also distinguish between two kinds of associations: 
type 1 and type 2.
Associations of type 1 include extended regions in the vicinity of groups, and 
could combine individual galaxies or groups.
On the other hand, associations of type 2 include only faint galaxies whose 
luminosities are not enough to reach the threshold to be defined as a group.
It is this latter kind of associations that interest us in the present project.

Almost twenty years later, \cite{Tully:2006} used the accurate estimator of 
the tip of the red giant branch distance achieved by the Hubble Telescope 
Advanced Camera for Surveys, to confirm five 
associations of dwarf galaxies previously identified, and to discover two more 
associations with similar properties.
They presented a detailed analysis of each association of dwarf galaxies and 
a description of their main dynamical properties.
The theoretical study of this kind of systems in the \mbox{$\Lambda$CDM} model
is the main goal of our work.

There are several studies about the interaction between 
dwarf galaxies, mainly between pairs of them.
\cite{Karachentsev:2008} presented a catalog of galaxy pairs in the near vicinity.
They focused on the large portion of binary systems compound by 
faint late--type dwarf galaxies.
They proposed that these binary systems formed by gas--rich dwarf galaxies may be 
at the stage prior to its merger. 
In \cite{Makarov:2012}, they reported a list of groups consisting of dwarf galaxies only.
Most of these systems reside in low-density regions and evolve unaffected 
by massive galaxies.
They compare the size, luminosity and the velocity dispersion 
of those groups with associations of dwarfs galaxies presented by \cite{Tully:2006}. 
They conclude that groups of dwarf galaxies have similar luminosity and 
velocity dispersion as associations of dwarf galaxies,
but the latter ones are undoubtedly larger than the groups.
Besides, their groups possess higher mass--to--luminosity 
ratios than the associations, which suggests that their systems 
have a larger amount of dark matter.

\begin{table*}
\centering
   \begin{tabular}{  c   c   c  c  c}
    \hline
    \hline
    \\
     ${\rm log}_{10}(M_{\rm max})$  & Number of systems & Number density & Maximum number of   & Total number of galaxies \\ 
     $[{\rm M}_{\odot}\,h^{-1}]$   & ($N_{\rm groups}$)   & [${\rm Mpc}^{-3}\,h^{3}$] & members per system & in systems \\ 
                       &         &         & ($N_{\rm members}$) & ($N_{\rm gals}$) \\ 
    \hline
    12.9 & 1508530 & 0.025 & 16380 & 15866313
    \\
    9.5 & 850856  & 0.014 & 42 & 3351797 
    \\
    9.0 & 606320 & 0.010 & 29 & 2256240 
    \\
    8.5 & 366713 & 0.006 & 19 & 1293617 \\
    \hline
    \hline
  \end{tabular}
  \caption{First column: Different stellar mass cuts used. Second column: Number of galaxy systems identified in each sample. 
  Third column: Number density of galaxy systems identified in each sample. Fourth column: Maximum number of members per system. 
  Fifth column: total number of semi-analytical galaxies belonging to systems for each sample.}
 \label{table:groups}
\end{table*}

In recent years, more studies have been presented about the interaction of 
dwarf galaxies.
For instance, \cite{Sales:2013} use observational data 
(Sloan Digital Sky Survey/Data Release 7, SDSS/DR7) to analyse the 
correlation between the abundance of satellite 
galaxies and the mass of the main galaxy.
They consider central galaxies in a wide range of stellar masses, 
including dwarf galaxies. 
They confirm that the amount of satellites galaxies grows 
as the satellite--to--primary stellar mass ratio increases when 
considering bright galaxies as centrals.
Otherwise, for the case of dwarf primaries, the amount of satellite galaxies 
does not depend on the mass of the primary.
They compare their results with mock catalogues, and conclude that they
are consistent with galaxy formation models in a $\Lambda$CDM Universe.
Later, \cite{Stierwalt:2017} presented seven spectroscopically confirmed groups 
compounded only by dwarf galaxies.
Even though each one of these groups has between three and 
five members, these groups are more compact and brighter than the 
associations presented by \cite{Tully:2006}.
\cite{Besla:2018} also use the SDSS observational data set to analyse isolated 
dwarf galaxies systems and compare their results with cosmological expectations.
They conclude that the majority of their isolated dwarf systems are pairs
(triplets are rare), and that more numerous systems are improbable from the 
cosmological point of view within the parameter range considered. 
They also argue that their results do not conflict with the groups presented by 
\cite{Stierwalt:2017} because most members of those groups do not satisfy their 
selection criteria.

The studies mentioned above were focused on the 
interaction between pairs of dwarf galaxies. 
Little is known about the associations of dwarf galaxies and their properties. 
Our main goal is to understand the existence and the properties of these 
rare objects within the standard $\Lambda$CDM cosmological model.
The existence of associations or groups of dwarf galaxies is critical 
to our understanding of structure formation at the low mass end 
of the stellar mass function - a regime 
known to constitute the small scale ``crisis'' of the current paradigm. 
The abundance of such objects will help us to understand group infall 
onto larger hosts, and galaxy conformity at the low mass end, 
as well as to shed light on the missing satellite problem. 
By quantifying the nature of dwarf associations, it is
plausible that the missing satellite problem becomes 
the ``missing dwarf association'' problem. 
Furthermore, given that dwarfs are the most abundant 
objects (by number) in the Universe, evaluating how much 
mass is locked up in these associations is important to 
understand the abundance of dark matter.
Moreover, a warm dark matter (WDM) model 
\citep{Lovell:2014} and the CDM scenario certainly predict different 
statistics for these objects.
So, the abundance of such groups could also be a probe of 
the nature of dark matter given the free streaming mass 
cut-off in 
a WDM model. 
It is thus important we understand the nature of these objects.

Although very few observational systems composed only by dwarf galaxies are 
currently known, this analysis is really important in the light of coming surveys, 
such as Dark Energy Spectroscopic Instrument (DESI, \citealt{DESI:2016}) 
and Legacy Survey of Space and Time (LSST, \citealt{LSST:2019}). 
Their precise measurements will produce the most detailed map of the 
nearby universe,
which will facilitate the discovery and analysis of these very particular systems.
Thus, in this project, we study these systems in a theoretical way by 
combining a cosmological dark matter only simulation with a semi--analytic model 
(SAM) of galaxy formation.
The low mass of the galaxies involved in this study calls for a 
simulation with good enough mass resolution.
To this aim, we use the $400\,h^{-1} \,{\rm Mpc}$
\textsc{Small MultiDark Planck} simulation (SMDPL) based on the Planck cosmology
\citep{Klypin:2016}, which is publicly available
in the \textsc{CosmoSim} database \footnote{\url{https://www.cosmosim.org}}. 
Dark matter haloes in this simulation are populated with galaxies generated by 
the semi--analytic model \sag~\citep[acronym for Semi-Analytic Galaxies,][]{Cora:2018}.
We compare our results with observational results presented by \cite{Tully:2006} 
and by \citet{Tully:2015}.
With such a comparison, we are able to analyse if cosmological simulations 
based on the $\Lambda$CDM paradigm are reliable to trace the hierarchical 
processes that are expected to influence the evolution of dwarf galaxies.

This paper is organized as follows.
We describe the SMDPL simulation and the most relevant
characteristics of the \sag~model in Section \ref{S_methods}. 
In Section~\ref{S_groups}, we define our systems, describe different samples
of galaxy ensembles and analyse their properties.
We then compare our results with observational results.
In Section \ref{S_haloes},
we analyse the dark matter haloes hosting galaxy members.
Finally, in Section \ref{S_conclusions}, we summarize our results and 
present our conclusions.

\section{Semi--analytical model} \label{S_methods}

\subsection{Dark matter simulation}
\label{sec:simSMDPL}

We use the SMDPL dark matter only simulation\footnote{doi:10.17876/cosmosim/smdpl/.},
which belongs to the series of MultiDark simulations with Planck cosmology. 
This simulation tracks
the evolution of $3840^3$ 
particles from redshift $z = 120$ to $z = 0$, within a periodic 
box of side--length of $400\,{\rm Mpc}\,h^{-1}$, 
achieving a mass resolution of $9.63\times10^7 \,{\rm M}_{\odot}\,h^{-1}$ 
per dark matter (DM) particle (see \cite{Klypin:2016} for further details).
SMDPL cosmological parameters are given by a flat $\Lambda$CDM model consistent with 
Planck measurements: \mbox{$\Omega_{\rm m}$ = 0.307115}, 
\mbox{$\Omega_{\rm B}$ = 0.048}, \mbox{$\Omega_{\Lambda}$ = 0.692885}, 
\mbox{$\sigma_{8}$ = 0.8228},
\mbox{$n_{\rm s}$ = 0.96} and \mbox{$h = 0.6777$}), \cite{Planck:2014}.

DM haloes have been identified with the \textsc{Rockstar} halo finder
\citep{Behroozi_rockstar} considering
overdensities with at least $N_\text{min}$~=~20~DM
particles. Merger trees have been constructed with
\textsc{ConsistentTrees} \citep{Behroozi_ctrees}.
The DM haloes can be detected over the background density 
(referred to as \textsl{main host} haloes) or lie within another 
DM haloes (subhaloes).
The virial mass of each main host halo is defined as the mass enclosed 
by a sphere of radius $r_{200}$, wihin which the mean density is a factor $\Delta=200$ times the critical density of the Universe 
$\rho_\textrm{c}$, i.e.,
\begin{equation}
   M_\textrm{200}(<r_\textrm{200}) = \Delta \rho_\textrm{c}
            \frac{4 \pi}{3} r_\textrm{200}^3.
   \label{eq:Mvir}
\end{equation}

The physical properties of subhaloes are estimated considering only the 
bound particles of the substructures identified by the halo finder.

Our SAM takes the information about the DM (sub)haloes extracted from the 
cosmological DM simulation and
their corresponding merger trees to generate the galaxy catalogue.

\subsection{ Semi-analytic model of galaxy formation SAG}

In this work, we use the latest version of the semi--analytic model 
SAG described in \cite{Cora:2018}. It is based on the SAM presented 
in \cite{Springel:2001} and has been further
modified and updated to include all main physical processes deemed to be 
important in galaxy formation. 
The combination of these processes regulate the circulation of mass and metals between 
the different baryonic components of the simulated galaxies (hot gas, cold gas, stars), 
determining their physical properties. 

Each new detected halo in the DM simulation hosts a galaxy generated by SAG,
so that main host haloes contain  central galaxies, while subhaloes are 
populated by satellite galaxies.
When the mass of the DM substructures becomes lower than the resolution limit 
of the halo finder, their satellites are referred to as orphans.
The positions and velocities of an orphan satellite are obtained from the numerical 
integration of its orbit, consistent with the potential well of the host halo. 
Following \cite{Gan:2010} and \cite{Kimm:2011}, the orbital evolution is affected 
by dynamical friction and tidal stripping. 
The assumed initial conditions are the position, velocity and virial mass of its 
subhalo the last time it was identified. 
When the halo-centric distance becomes smaller than $10$ per cent of the 
virial radius of the main host halo, the orphan satellite is considered 
to be merged with the corresponding central galaxy.

As a result of the hierarchical growth of structure, the virial mass of dark 
matter haloes changes between consecutive outputs of the simulation. 
This change determines the increase of the mass of hot gas of central galaxies, 
which is defined as the baryon fraction of the virial mass of the main host 
halo at each snapshot of the simulation.
Galaxies keep their hot gas halo when they become satellites, but it is gradually 
removed by the action of tidal stripping and ram pressure stripping.
Thus, gas cooling of the hot gas phase takes place in both central
and satellite galaxies \citep{Cora:2018}, replenishing their cold gas discs 
that have been formed by the gas inflows.
This baryonic component is also affectd by ram pressure 
stripping \citep{Tecce:2010}, when a significant fraction (90 per cent) 
of the hot halo is removed.
Stars form in both quiescent and bursty modes. 
The former takes place when the mass of the cold gas disc exceeds a critical limit, 
while the latter is triggered by both mergers and disc
instabilities, giving place to the formation of a stellar bulge
where the transferred gas is gradually consumed  \citep{Gargiulo:2015}.
Bulge formation is accompanied by the growth of a super--massive
black hole, which also increases its mass during gas cooling
producing an active galactic nucleus (AGN) with its consequent feedback \citep{Lagos:2008}.
Both type Ia and II supernovae (SNe) produce energetic feedback. 
Together with stellar winds, they contaminate all baryonic components with 
different chemical elements. The processes involved are recycling of mass 
and metals that had been locked in stars, reheating of cold gas,
ejection of the hot gas out of the halo, and reincorporation of this ejected 
gas \citep{Cora:2006, Collacchioni:2018}; timescales for mass loss and lifetime 
of SNe progenitors are taken into account. 

The different physical processes are regulated by free efficiencies and 
parameters that have been adjusted by comparing the model results against 
a given set of observed galaxy properties.
The set of best-fitting values for the free 
parameters of SAG for the SMDPL simulation have been obtained 
by the Particle Swarm Optimization 
\citep[PSO,][]{Ruiz:2015}, using as observational constraints the same 
data sets used in previous 
calibrations of the SAG model: the stellar mass functions at $z = 0$ and $z = 2$, 
the star formation rate distribution function at $z=0.14$, the fraction of 
cold gas as a function of stellar mass and the bulge mass-black hole mass 
relation (see \citet{Knebe:2018} for a complete description of these data sets).
It is worth noticing that the values of the free parameters obtained for the 
SMDPL simulation differ from those obtained for the lower resolution MDPL2 
simulation \citep{Klypin:2016}, for the same version of the code. 
This arises because the parameter that regulates the redshift dependence 
of the SNe feedback was allowed to vary when  calibrating the MDPL2 
simulation while it was fixed for the SMDPL simulation. 
This choice was based on the better results obtained for the fraction 
of $z=0$ quenched galaxies when such a parameter was lowered with 
respect to the one obtained in the MDPL2 calibration 
\citep[see][for  a detailed discussion]{Cora:2018}. 

In order to validate the results obtained from applying the 
semi--analytic model SAG to the SMDPL simulation, we analyze 
the stellar mass of semi--analytical galaxies, the virial mass 
of their haloes and the relation between them. 
The bottom left panel of Figure \ref{fig:SAG} shows the median 
stellar--virial mass relation for all semi--analytical galaxies.
Present--day halo masses are considered for central galaxies, 
while infall halo masses for satellites. 
Vertical error bars correspond to $25$ and $75$ per cent quartiles.
We compare our results with the correlation obtained from the
abundance matching technique presented by \cite{Moster:2013} (grey dashed line)
and find a very good agreement at the low-mass end,
which is of particular interest for this work. 
The bottom right panel of Figure \ref{fig:SAG} displays the stellar mass 
function for all semi--analytical galaxies in the volume of the SMDPL 
simulation (black solid line). 
Our results are compared with observational data compiled by 
\cite{Henriques:2013} \citep{Li:2009, Baldry:2008} (grey filled circles), 
and they show a rather good agreement in the low-mass end.
The excess of galaxies at the high-mass end has already been noted and discussed 
by \citet{Cora:2018}; they do not affect the results of the current work which 
is focused on a much lower stellar mass range. 
The top left panel of Figure \ref{fig:SAG} displays the virial mass 
function of the (sub)haloes
hosting these semi--analytical galaxies (black solid line).
Coloured arrows indicate different stellar mass thresholds, 
$M_{\rm max}$, used to define three different samples of dwarf galaxies 
systems as it is described in sub--section \ref{sample}.

\begin{figure}
  \centering
  \includegraphics[width=0.48\textwidth\hspace{0pt}]{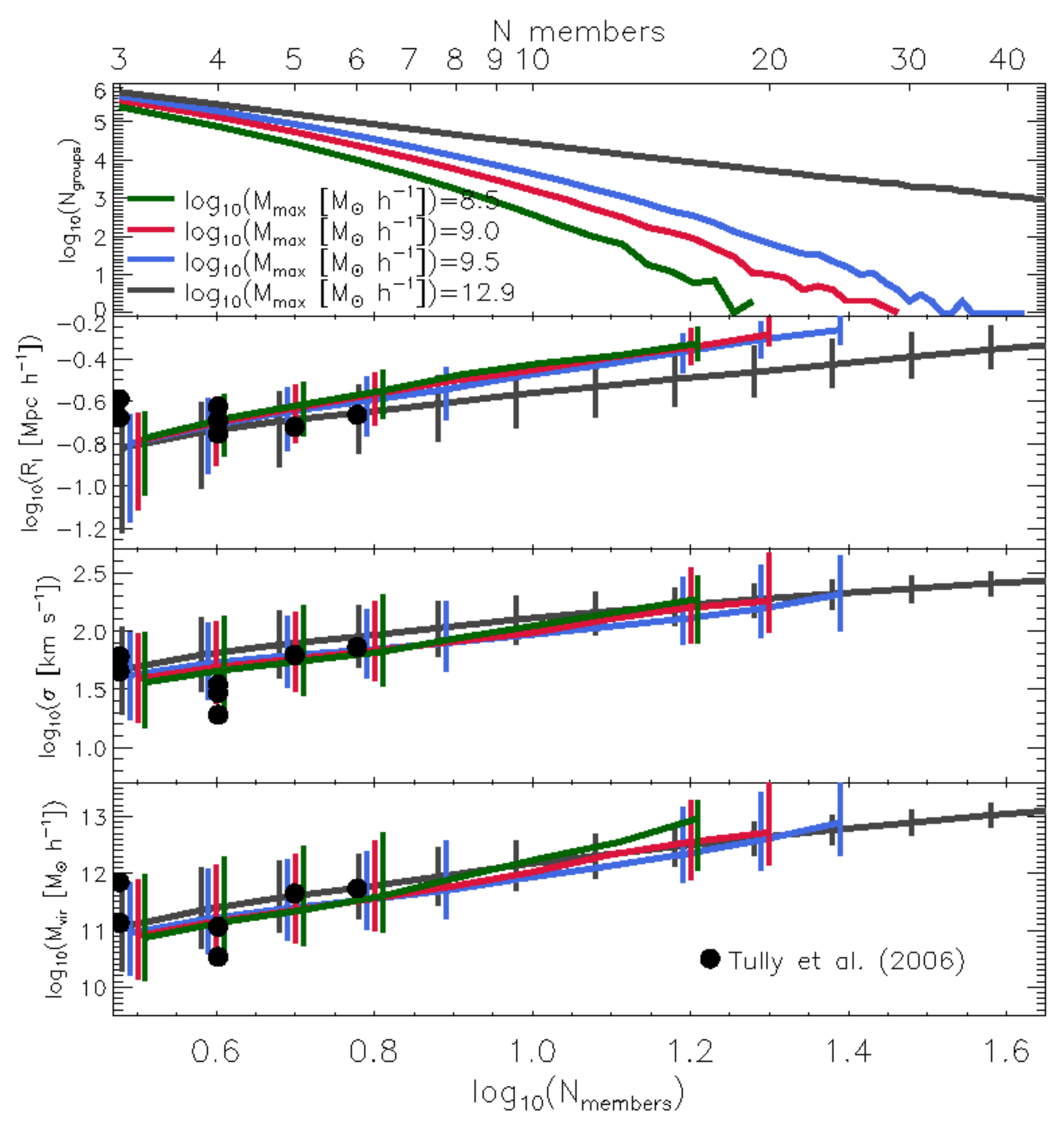}
  \caption{Galaxy system properties as a function of the number of members per system, 
  for our three samples of dwarf galaxy systems (green, red, blue lines) and the 
  complete sample (grey line).
  From top panel to bottom panels: Number of systems, median values of inertial 
  radius, median values of velocity dispersion and median values of virial mass.
  Median values are shown only if there are $10$ or more objects for a given number 
  of members. Vertical error bars correspond to 1$\sigma$ errors. 
  Black filled circles show observational associations of dwarf galaxies 
  presented by \citep[]{Tully:2006}.}
  \label{fig:properties}
\end{figure}

\section{Systems of Dwarf Galaxies} \label{S_groups}

\subsection{Systems sample} \label{sample}

From the galaxy population generated by applying the model SAG to the SMDPL 
simulation, we extract our sample of semi-analytical galaxies. 
To restrict our sample to well-resolved objects, we impose a condition 
to stellar mass,  $M_{\rm *} \ge 10^{6.8}\,{\rm M}_{\odot}\,h^{-1}$,
and virial mass, $M_{200} \ge 10^{9.28} \, {\rm M}_{\odot}\,h^{-1}$ 
(equivalent to $20$ DM particles).
Our sample has a total of $26,506,948$ semi--analytical galaxies, with stellar masses 
ranging between 
$6.8 < {\rm log_{10}}(M_{\rm *}[{\rm M}_{\odot}\,h^{-1}]) < 12.9$
and virial masses ranging between 
$9.28 < {\rm log_{10}}(M_{200}[{\rm M}_{\odot}\,h^{-1}]) < 15.17$.

From this sample of semi-analytical galaxies,
we identify systems of galaxies, having a minimum of 3 members,
using a {\em friends-of-friends} (FoF) algorithm \citep{H&G:1982} 
with a linking length of $0.4 \,{\rm Mpc}\,h^{-1}$.
This value was calibrated to reproduce the characteristic 
sizes and the virial masses of the observational sample of associations 
of dwarf galaxies presented by \cite{Tully:2006}.
To select this value, we vary the linking length parameter between 
$0.3 \,{\rm Mpc}\,h^{-1}$ and $0.5 \,{\rm Mpc}\,h^{-1}$. 
Bottom left panel of Figure 3 shows the virial mass as a function 
of the inertial radius (see next Subsection for the definition of these 
properties).
Inner contour levels enclose 25 per cent of each sample for 
three different linking lengths: $b=0.3\,{\rm Mpc}\,h^{-1}$ 
(short dashed blue line), $0.4\,{\rm Mpc}\,h^{-1}$ 
(solid blue line) and $0.5\,{\rm Mpc}\,h^{-1}$ (long dashed blue line).
Filled circles show the 7 observational associations of dwarf galaxies 
taken from \cite{Tully:2006}.
The solid line ($b=0.4 \,{\rm Mpc}\,h^{-1}$) encloses
5 out of 7 observed associations, while the others, 
short dashed line ($b=0.3\,{\rm Mpc}\,h^{-1}$) 
and long dashed line ($b=0.5\,{\rm Mpc}\,h^{-1}$), only two. 
So, comparing with the observational results presented by \cite{Tully:2006}, 
we found that $0.4 \,{\rm Mpc}\,h^{-1}$ is the best choice for the analysis 
of these associations, without the need for a fine tuning. 
We analyse dwarf galaxies associations at redshift $z=0$ due to the fact 
that the only associations of this type, detected so far, are at 
redshift $z=0$.

In order to build up our sample of FoF objects composed exclusively 
by dwarf galaxies, we remove those systems that have one (or more) 
galaxy member more massive than a stellar mass threshold $M_{\rm max}$.
Although there is no a unique definition for dwarf galaxy, many works 
adopt a maximum stellar mass value close to 
\mbox{${\rm log}_{10}(M_{\rm max}[{\rm M}_{\odot}\,h^{-1}]) = 9.0$} 
(see for example, \cite{Fattahi:2013}, \cite{Stierwalt:2017} and 
\cite{Besla:2018}, among others).
To explore the sensitivity of our results to this specific stellar mass cut, 
our analysis was preformed analysing three samples with 
\mbox{${\rm log}_{10}(M_{\rm max}[{\rm M}_{\odot}\,h^{-1}]) = 8.5$}, $9,0$ and $9.5$.
Table (\ref{table:groups}) summarizes the main information of our four samples, namely 
the full sample and the three sub samples where upper mass cuts have been applied to 
the total FoF stellar mass.
Columns show the stellar mass cut used, the number of systems, the number density 
of systems, the maximum number of members per system and the total number of 
semi--analytical galaxies belonging to systems.

\begin{figure}
  \centering
  \includegraphics[width=0.48\textwidth\hspace{0pt}\vspace{0pt}]{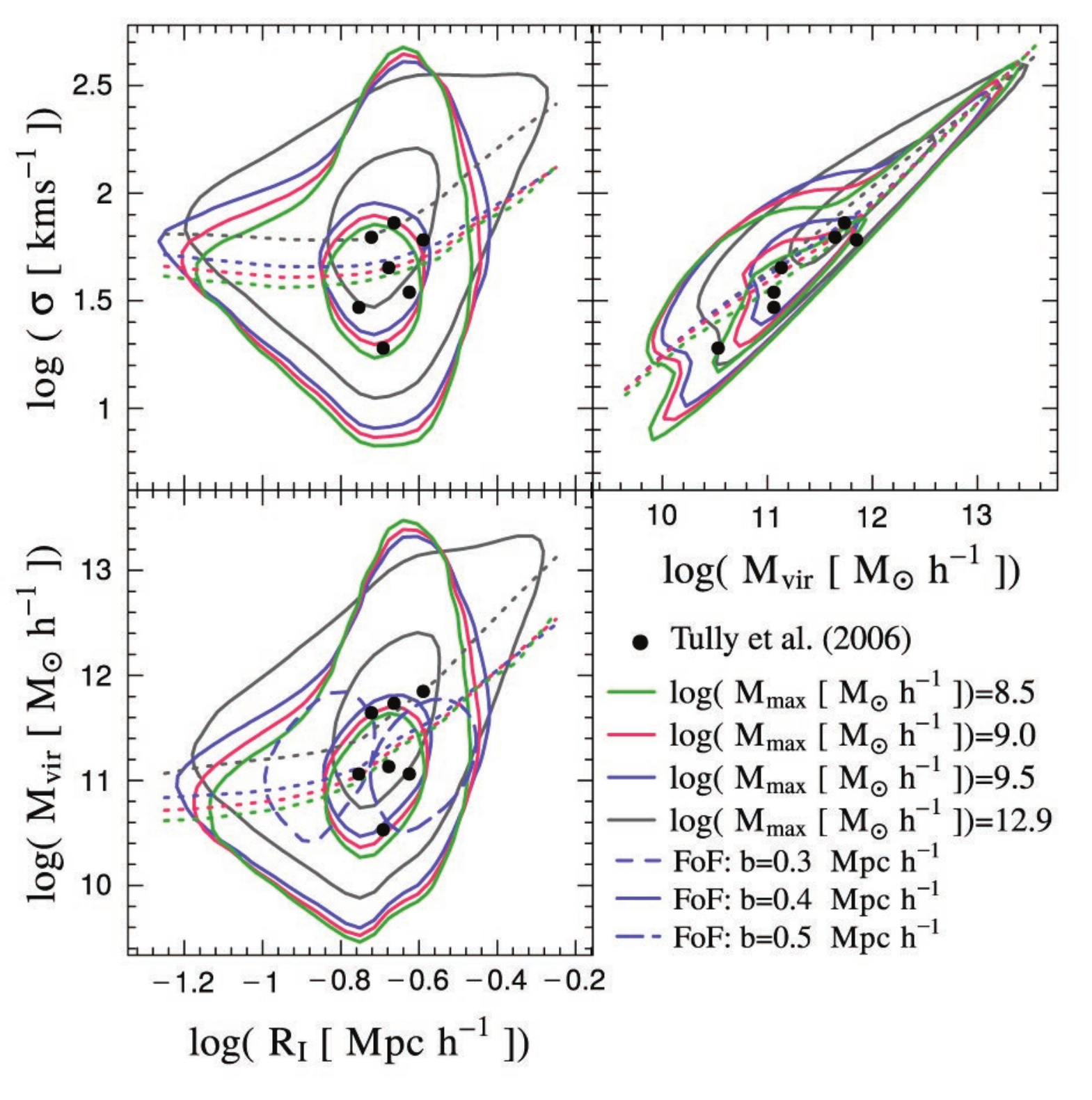}
  \caption{Scaling relations between size, velocity dispersion and mass for 
  our three samples of dwarf galaxy systems (solid green, red and blue 
  lines, see labels)
  and the complete sample (solid grey lines).
   They are compared with observational results for 7 dwarf galaxy associations taken from
   \citet[][black filled circles]{Tully:2006}. 
   Solid inner (outer) lines show contour levels enclosing $25$ ($75$) per cent.
   Dotted lines show median values computed using bins in the horizontal axis of 
   constant width of $0.1$ in logarithmic scale. 
   Contour levels indicated by dashed blue lines in the bottom 
   left panel enclose $25$ 
   per cent of the sample, for two different linking lengths:  $b=0.3\,{\rm Mpc}\,h^{-1}$ (short 
   dashed blue line) and $0.5\,{\rm Mpc}\,h^{-1}$ (long dashed blue line). 
   }
  \label{fig:scaling_rel}
\end{figure}

\begin{figure}
  \centering
  \includegraphics[width=0.48\textwidth\hspace{0pt}\vspace{0pt}]{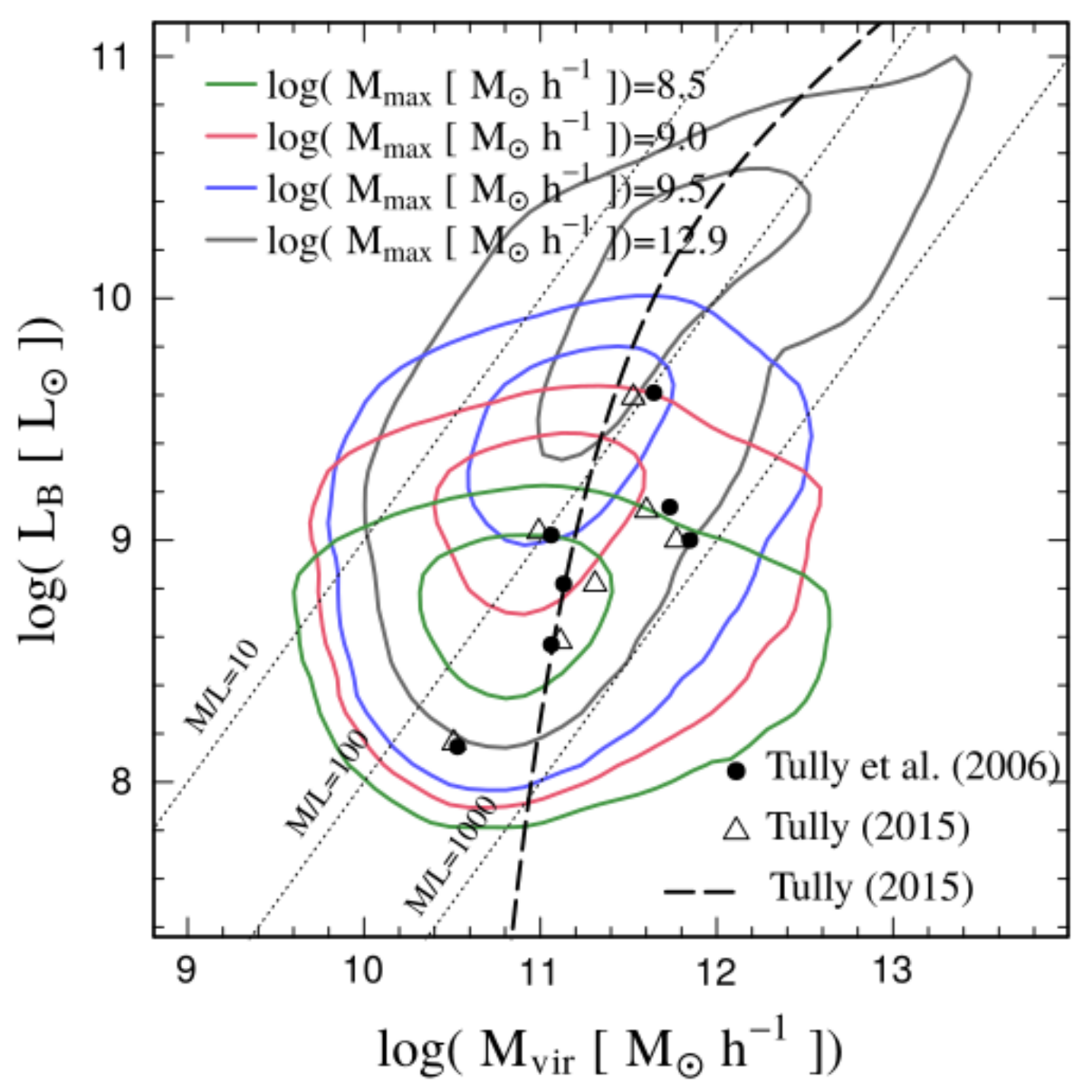}
  \caption{Luminosity in the blue band as a function of the virial mass 
  for our three samples of dwarf galaxy systems (green, red and blue lines) 
  and the complete sample (grey lines). 
  Solid inner (outer) lines show contour levels enclosing $25$ ($75$) per cent.
  They are compared with observational results for 7 dwarf galaxy associations 
  taken from \citet[][black filled circles]{Tully:2006} and 
  \citet[][open triangles]{Tully:2015}.
   The black dashed line shows the empirical fit to the data presented by \citet{Tully:2015}.
   Dotted lines correspond to three different values of the mass--to--light relation (M/L) 
   as it is indicated.}
  \label{fig:mvir_lum}
\end{figure}

\subsection{Systems properties} \label{S_properties}

By resolving stars at the tip of the red giant branch using the Hubble Space Telescope 
Advance Camera for Surveys, accurate distances were determined by \cite{Tully:2006} 
to seven associations of dwarf galaxies.
Five of these were previously found by \cite{Tully:1987} and two more new 
associations were presented.
Although this is one of the oldest observational sets, it is still 
nowadays the most complete and homogeneous observational sample 
of data of the associations of dwarf galaxies.
\citet{Tully:2015} and \cite{Kourkchi:2017} also study these associations but they 
do not present a complete and updated compilation of all their main dynamical 
properties.
Only the luminosities and virial masses were re-estimated in \citet{Tully:2015} 
for the seven associations.
So, we compare the intrinsic properties of the observational 
associations presented by \cite{Tully:2006}, and with \cite{Tully:2015}
when possible, with our systems identified in the 
\mbox{$\Lambda$CDM} model.
With this aim, we follow the procedure presented by 
\cite{Tully:2006} to compute the main intrinsic properties of 
our systems: inertial radius ($R_{\rm I}$) as an indicator of the size 
of the system, the velocity dispersion ($\sigma$) and 
the virial mass ($M_{\rm vir}$).

The inertial radius is defined as

\begin{equation}
R_{\rm I} = \left( \sum_{i}^{N} r_{i}^{2}/N \right) ^{1/2}.
\label{eq:radius}
\end{equation}

\noindent where $r_{i}$ is the three--dimensional distance of a galaxy from the 
system centroid, and the sum for each system is performed over all members ($N$). 
The velocity dispersion is 

\begin{equation}
\sigma =\left [\sum_{i}^{N} v_{i}^{2}/(N-1)\right]^{1/2}.
\label{eq:vel}
\end{equation}

\noindent where $v_i$ is the one--dimensional velocity difference between a 
galaxy and the system mean.
To estimate the virial mass of the system, we use the expression adopted by 
\cite{Tully:2005} and \cite{Tully:2006}, that is

\begin{equation}
M_{\rm vir} = 3[(N-1)/N]\sigma^{2}R_{\rm G}/G. 
\label{eq:mvir}
\end{equation}

\noindent where the radius is given by 
\mbox{$R_{\rm G}=N^2/\sum_{\rm pairs}(1/r_{ij})$}, 
where $r_{ij}$ is the separation between pairs in the system counted only once.
It is worth noticing that this equation assumes systems in virial 
equilibrium, a hypothesis that is probably far from being true for both 
observational and theoretical associations. 

The upper panel of Figure \ref{fig:properties} shows the number of systems as a 
function of the number of members for our three samples of dwarf galaxy 
systems and the complete sample, colour-coded as indicated in the legend.
The number of systems decreases as the number of members increases, as expected.
Notice that our systems of dwarf galaxies are largely (more than 75 per cent) 
dominated by systems made of 3 or 4 members.
These power law distributions cover a range of slopes from $\sim -2.6$ 
for the complete sample (grey line) to $\sim -5.4$ for the sample 
\mbox{${\rm log}_{10}(M_{\rm max}[{\rm M}_{\odot}\,h^{-1}]) = 8.5$} 
(green line). 
It means that dwarf galaxy associations show a relative deficit of 
numerous member systems.
Second, third and fourth panels show the median values of inertial radius 
($R_{\rm I}$), velocity dispersion ($\sigma$) and virial mass ($M_{\rm vir}$) 
as a function of the number of members of the system, respectively.
Median values are shown only if there are $10$ or more objects for a given 
number of members.
Systems with more members are bigger, with higher velocity dispersion 
and more massive than systems with fewer members.
Over plotted on the bottom three panels are the observational data from \cite{Tully:2006}.
Note that, in general, the data are within one $\sigma$ of our predicted values.

\subsection{Comparison with observational results} \label{S_obs}

Figure \ref{fig:scaling_rel} compares the scaling relations between 
inertial radius ($R_{\rm I}$), velocity dispersion ($\sigma$) and 
virial mass ($M_{\rm vir}$) of the systems of dwarf galaxies 
predicted by our model and the observational results presented by 
\cite{Tully:2006}.
Solid inner (outer) lines show contour levels enclosing 
$25$ ($75$) per cent of each sample, for our three samples of dwarf 
galaxy systems (green, red and blue lines) and the complete sample (grey lines).
Dotted lines show median values computed using bins in the horizontal 
axis of constant width of $0.1$ in logarithm scale.
Black filled circles show observational associations of dwarf galaxies 
presented by \cite{Tully:2006}.
This figure shows that calibrating the linking length to reproduce 
reasonably well the characteristic size of associations of dwarf 
galaxies, allows us to find also good agreement in the mass and the 
velocity dispersion without any extra tuning in our model.
In each one of the scale relationships, six of the seven observational
associations of dwarf galaxies are within contour levels enclosing 
$25$ per cent of each sample.
In spite of the significant scatter, from the median values of each property 
and the contour levels, we can conclude that in the $\Lambda$CDM model we 
find dwarf galaxy systems which properties are comparable with 
observable dwarf galaxy associations. 

In addition to the dynamical properties, it is important to study 
the luminosity of these systems. 
The mass--to--light ratio (M/L) gives information about the true nature 
of these systems.
We analyse the relationship between the {\em B}--band 
luminosity ($L_{\rm B}$) and the virial mass ($M_{\rm vir}$) of our 
dwarf galaxy systems. 
Figure \ref{fig:mvir_lum} shows the {\em B}--band luminosity as a function of 
virial mass for our three samples of dwarf 
galaxy systems (green, red and blue lines) and the complete sample (grey lines).
Solid inner (outer) lines show contour levels enclosing 
$25$ ($75$) per cent of each sample.
These results are compared with observations for the seven dwarf 
galaxy associations taken from \cite{Tully:2006} (black filled dots) and 
\cite{Tully:2015} (empty triangles).
\cite{Tully:2015} analyses the associations of dwarf galaxies presented 
by \cite{Tully:2006}, and re-estimates their luminosities and virial masses.
Black dotted line shows the empirical fit to the data presented by equation 16 
in \cite{Tully:2015}.
Luminosities are affected by intrinsic dust attenuation in both the observations and the model. In the latter,
the attenuation is added using the calculation presented by \cite{DeLucia:2004}, as explained in \cite{Cora:2006}.
We find a reasonable agreement between the luminosity--mass relation of our 
semi--analytical systems of dwarf galaxies and the observational results.
In this case, all the observational associations of dwarf galaxies 
are within contour levels enclosing $75$ per cent of each sample.
Furthermore, it seems that the mass--to--light ratio is relatively high 
if these systems are bound systems.
From Figures \ref{fig:scaling_rel} and \ref{fig:mvir_lum}, we also infer 
that the highest stellar mass threshold used in our analysis, i.e., 
\mbox{${\rm log}_{10}(M_{\rm max}[M_{\odot}\,h^{-1}]) = 9.5$}, 
is the one which best matches the observational results.

\begin{figure}
  \includegraphics[width=0.48\textwidth\hspace{0pt}]{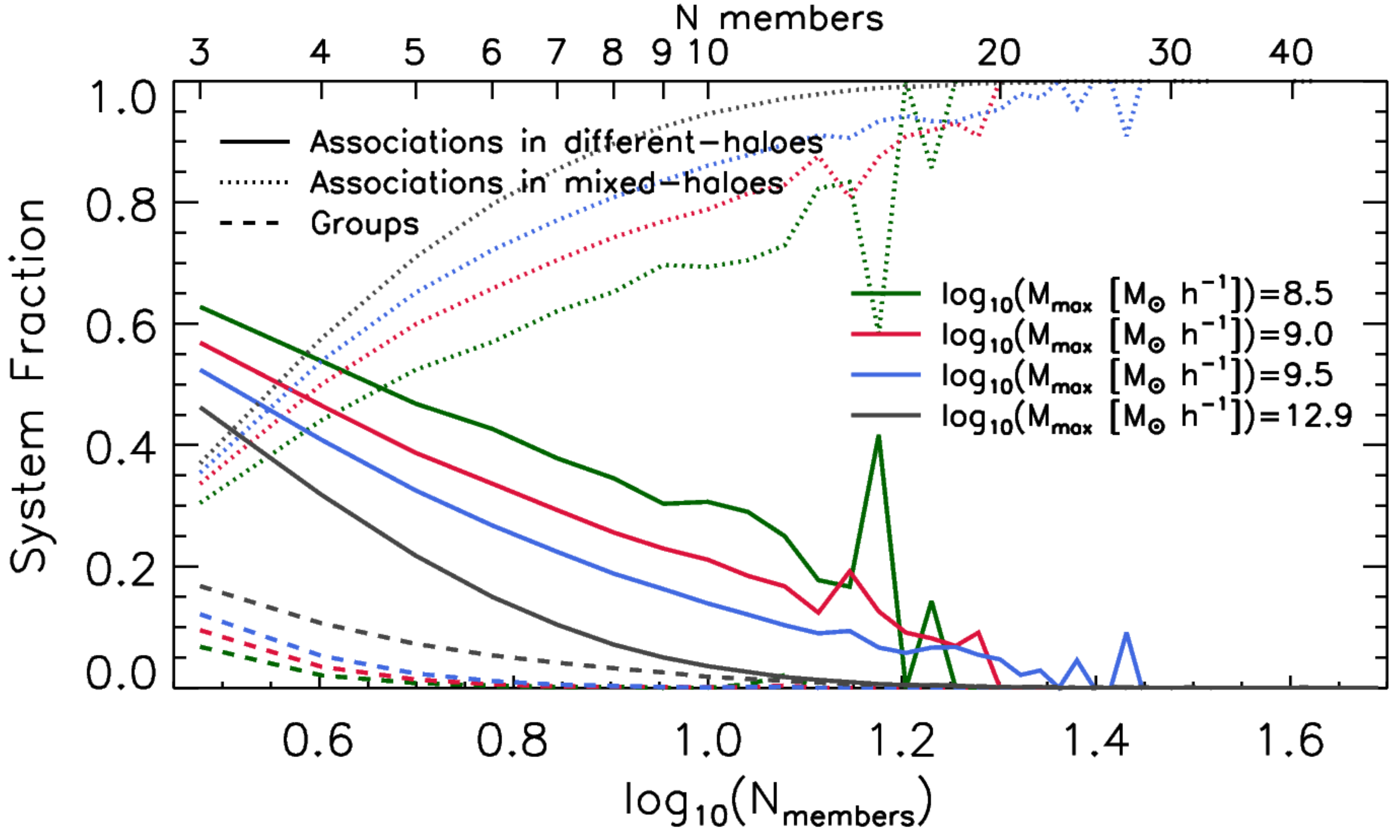}
  \caption{Fraction of {\em Associations in different--haloes} 
  (solid lines), {\em Associations in mixed--haloes} (dotted lines) and 
  {\em Groups} (dashed lines), as a function of the number of 
  members per system, for our three samples of dwarf galaxy systems 
  (green, red, blue lines) and the complete sample (grey lines).}
  \label{fig:fraction}
\end{figure}

\begin{figure*}
    \begin{subfigure}{
        \includegraphics[width=0.34\textwidth]{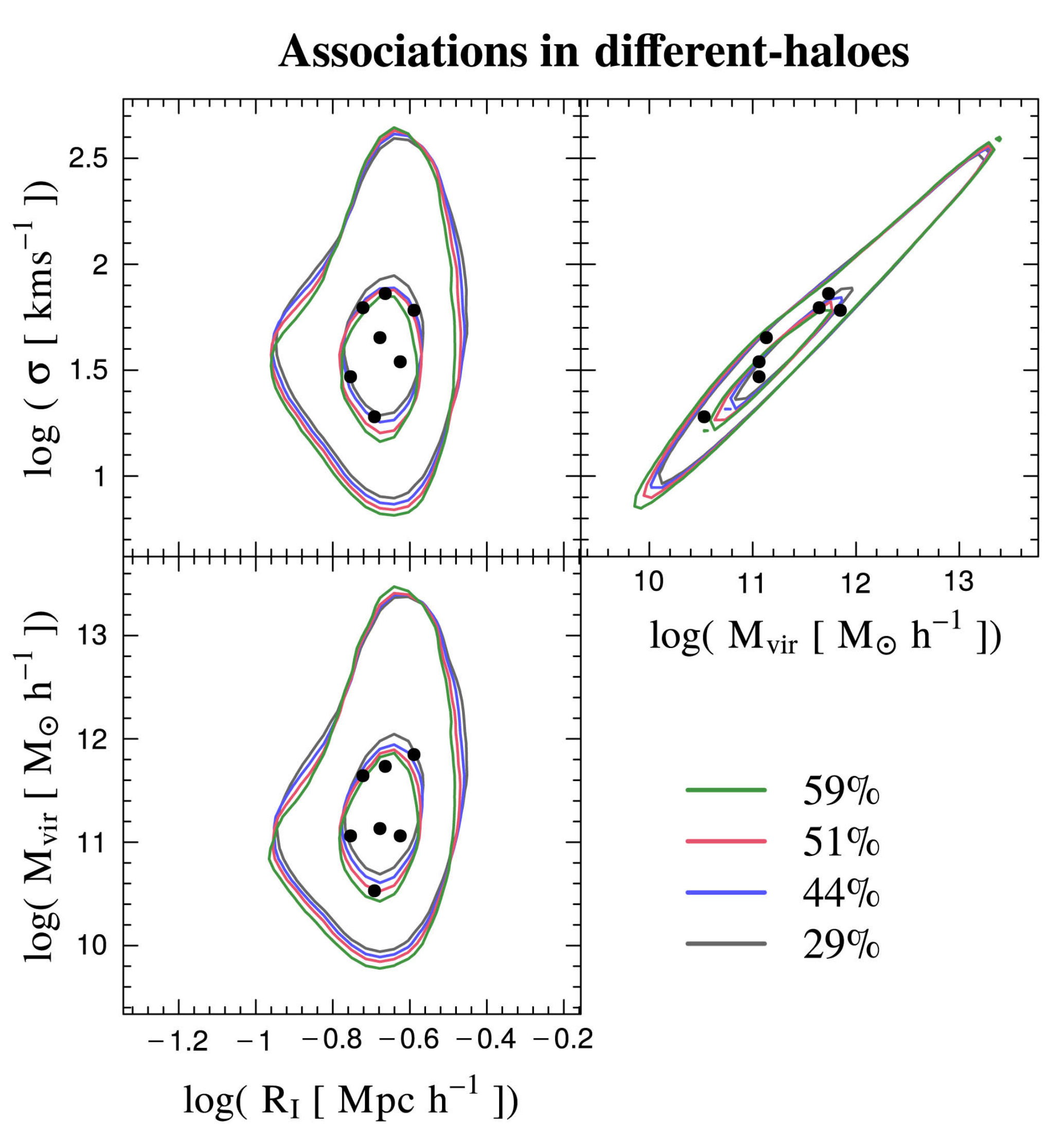}
        \includegraphics[width=0.325\textwidth]{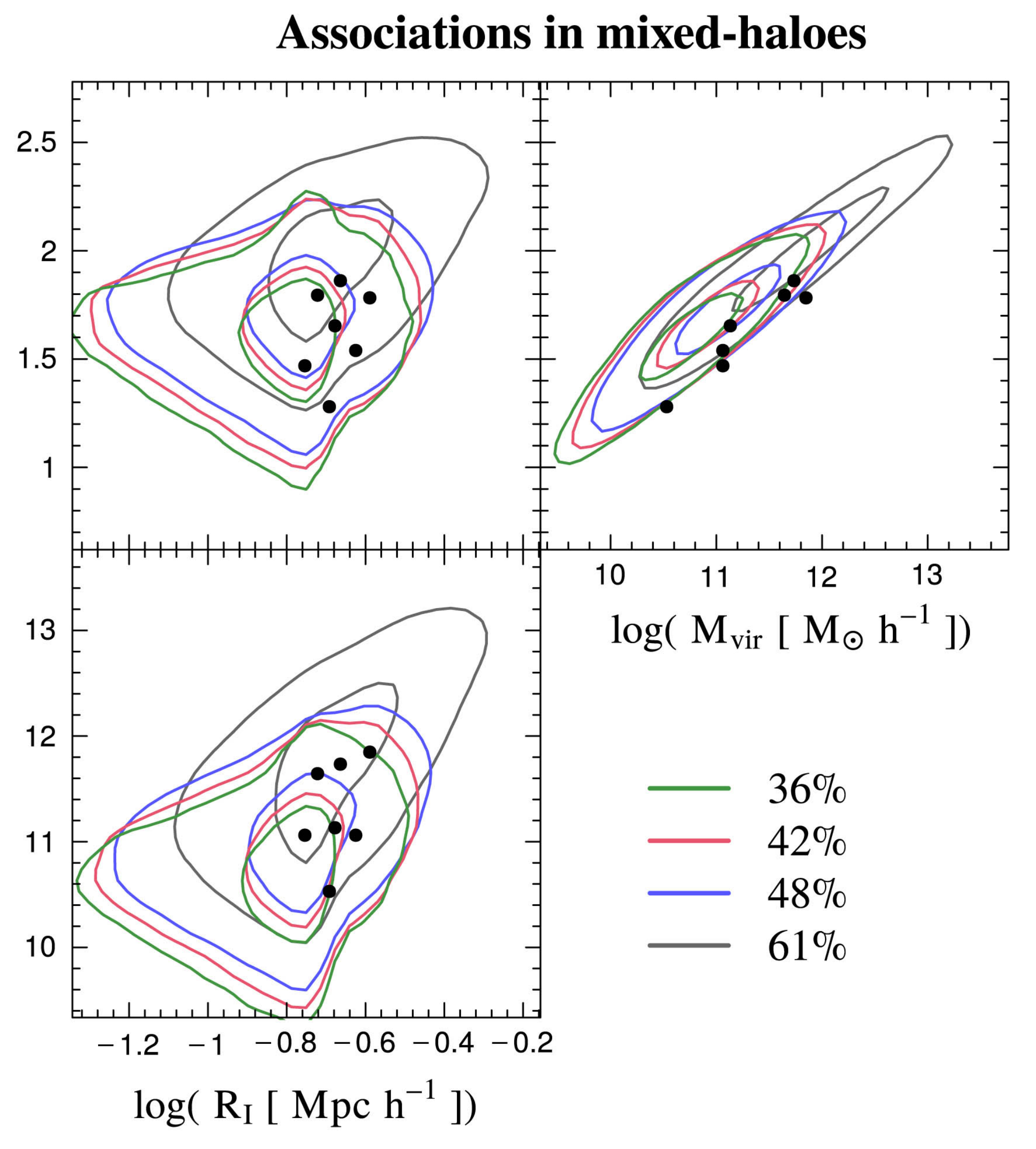}
        \includegraphics[width=0.324\textwidth]{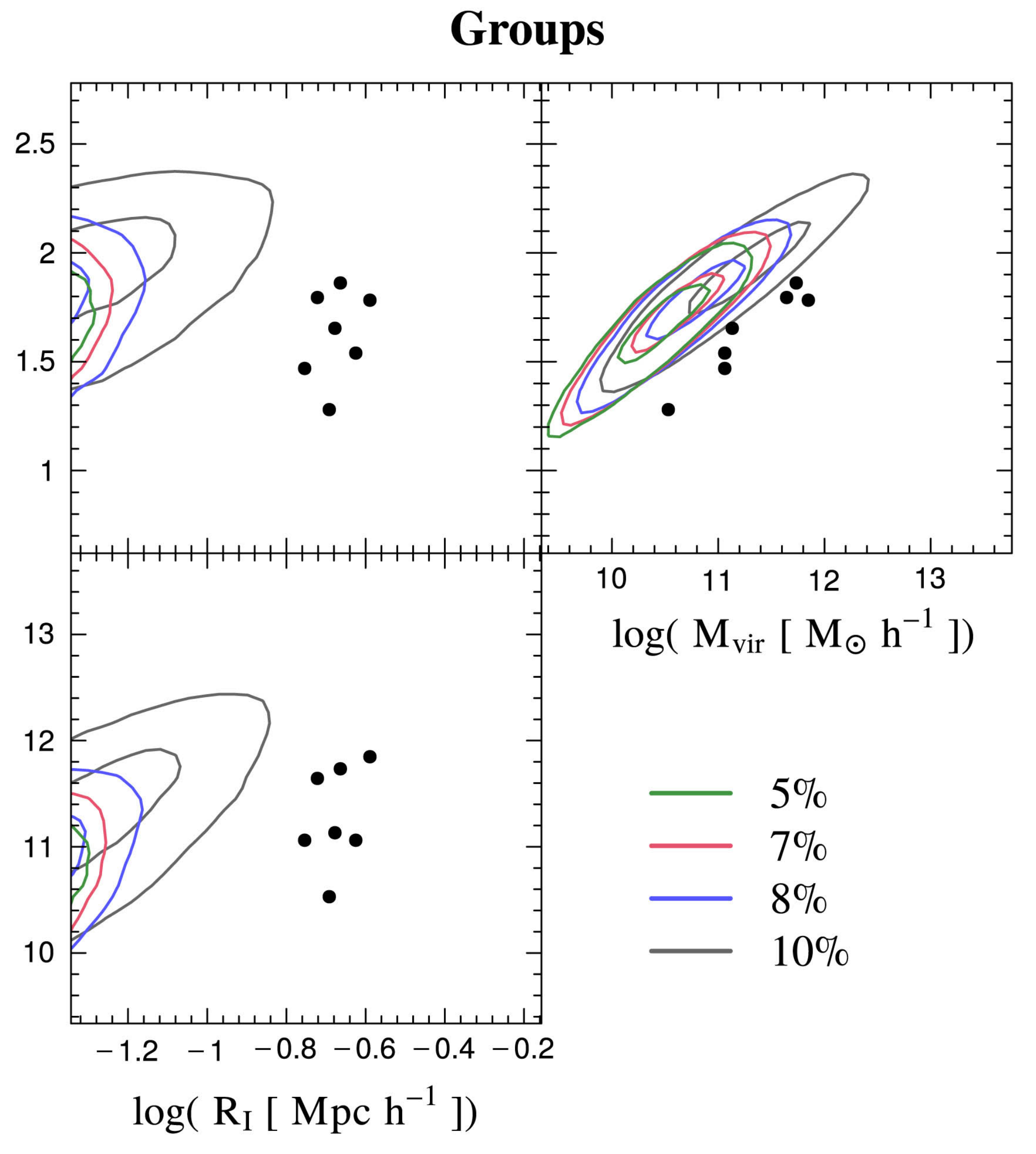}}
    \end{subfigure}
    \caption{Scaling relations between size, velocity and mass for our three 
    samples of dwarf galaxy systems (green, red, blue lines) and the 
    complete sample (grey lines),
    which are splitted in three different sub--samples defined according to the
    belonging of their galaxy members to main host haloes:
    {\em Associations in different--haloes} (left panels), {\em Associations in mixed--haloes} 
    (middle panels) and {\em Groups} (right panels).
    Solid inner (outer) lines show contour levels enclosing $25$ ($75$) per cent.
    They are compared with observational results for the seven dwarf galaxy associations taken from
    \citep[]{Tully:2006} (black filled circles).
    For each sample of systems characterized by one of the $M_{\rm max}$ thresholds adopted, 
    it is specified the percentage of systems in each sub--sample.
    Colour code is the same as in previous figures.}
    \label{fig:change}
\end{figure*}

\section{Nature of systems of dwarf galaxies} \label{S_haloes}

In the previous section, we infer that both our systems and the associations 
presented by \cite{Tully:2006}, present \mbox{M/L $>$ 100}, so if the associations 
are bound then their M/L values are very high.
To shed some light on this subject, we analyse the assignment of dwarf 
galaxy members of systems to dark matter haloes.
The main question is whether dwarf galaxies which are members of these kind of systems 
live in the same dark matter halo or in different dark matter haloes.
In that sense the reader will note that \cite{Kourkchi:2017} propose a definition of 
``associations'' for systems that belong to different dark matter haloes, while 
they define as ``groups'' those systems who live in the same dark matter halo. 
They used these definitions to classify different kind of systems, 
not only those consisting of dwarf galaxies.
In this scheme, characteristic sizes of associations correspond to the 
first turnaround radius, while the characteristic sizes of groups 
correspond to the second turnaround radius.

Specifically, we identify if the  members of each  of our systems 
are located in the same main host dark matter halo or if they are located 
in different dark matter haloes, using information given by 
\textsc{Rockstar} halo finder.
For this purpose, we define three different sets for each one of our 
semi--analytical samples according to the following conditions: 
(i) systems with all their galaxy members belonging to the same halo; 
(ii)  systems with all their galaxy members belonging to different main host dark matter haloes; 
(iii) a mixed of the two, i.e., systems for 
which some of the galaxies belong to the same dark matter halo, but others 
belong to different host haloes. 
According to the classification of \cite{Kourkchi:2017}, the first classification would be 
considered as ``groups''.
Hereafter, we will refer to each of these sub--samples as: 
(i) {\em Groups}, (ii) {\em Associations in different--haloes}, and (iii) {\em Associations in mixed--haloes}.

Figure \ref{fig:fraction} shows the fraction of 
{\em Associations in different--haloes} (solid lines), 
{\em Associations in mixed--haloes} (dotted lines) and 
{\em Groups} (dashed lines), as a 
function of the number of members per system, 
for our three samples of dwarf galaxy systems
(green, red, blue lines) and the complete sample (grey lines).
It is apparent that most of the systems belong to {\em Associations in different--haloes} and 
{\em Associations in mixed--haloes} sub--samples, while only about 10 per cent (or less) 
belong to the {\em Groups} sub--sample.
Although it is a function which depends on the number of members, 
on average, about 50 per cent of our systems of dwarf galaxies reside 
in {\em Associations in different--haloes}, about 40 per cent in 
{\em Associations in mixed--haloes} and less than 10 per cent in 
{\em Groups}. 
As the number of members increases, the fraction of dwarf galaxy systems that
belong to {\em Associations in mixed--haloes} becomes increasingly important.

Next, we analyse the differences in the dynamical properties of the systems 
depending on whether their galaxy members are located in the same host 
dark matter halo or in different ones.
Figure \ref{fig:change} shows the scaling relations presented in Figure 
\ref{fig:scaling_rel} but now subdivided into the three samples of 
dwarf galaxy systems (green, red, blue lines) and the complete sample 
(grey lines), and split into {\em Associations in different--haloes} (left panels), 
{\em Associations in mixed-haloes} (middle panels) and in {\em Groups} (right panels).
Solid inner (outer) lines show contour levels enclosing 
$25$ ($75$) per cent of each sample.
Bottom right labels show the percentage of each sample. 
For comparison, we also plot the observational results for the seven dwarf 
galaxy associations taken from \cite{Tully:2006} (black filled circles), as in 
Figure \ref{fig:scaling_rel}.
As expected, the {\em Associations in different--haloes} 
and {\em Associations in mixed--haloes} reproduce much better the characteristic 
sizes of the observational sample of dwarf galaxies associations.
In the {\em Associations in different--haloes} case, the seven observational 
associations of dwarf galaxies are within contour levels enclosing $25$ 
per cent of each sample if the scaling relation includes the inertial radius. 
In the case of the scaling relation between velocity and mass, the seven 
observational associations of dwarf galaxies are within contour levels 
enclosing $75$ per cent of each sample.
Conversely, {\em Groups} are systematically smaller by a factor of $\sim 5$, 
and none of the seven observational associations of dwarf galaxies are 
within their contour levels for any of the samples.

\begin{figure*}
    \begin{subfigure}{
        \includegraphics[width=0.34\textwidth]{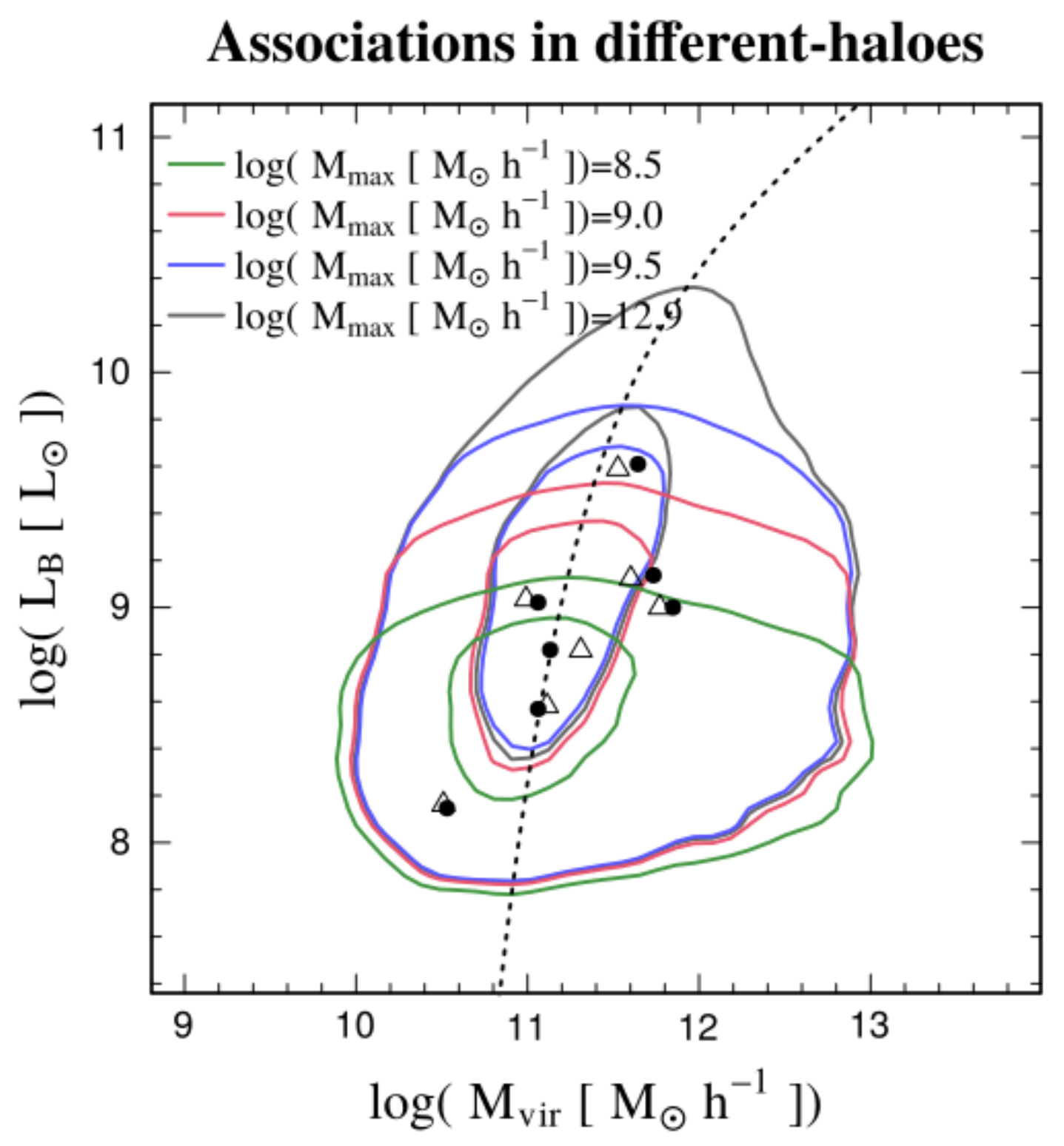}
        \includegraphics[width=0.32\textwidth]{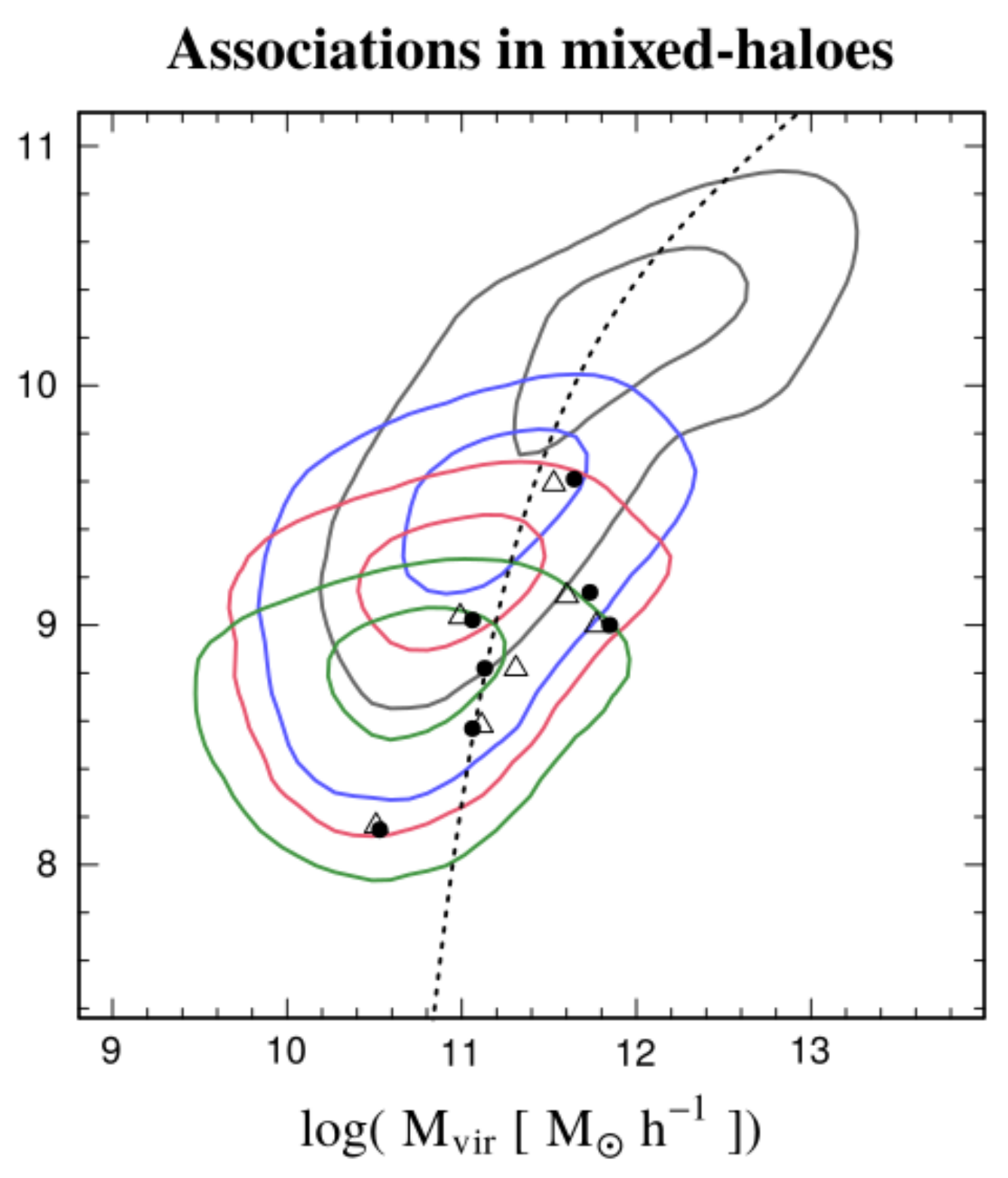}
        \includegraphics[width=0.32\textwidth]{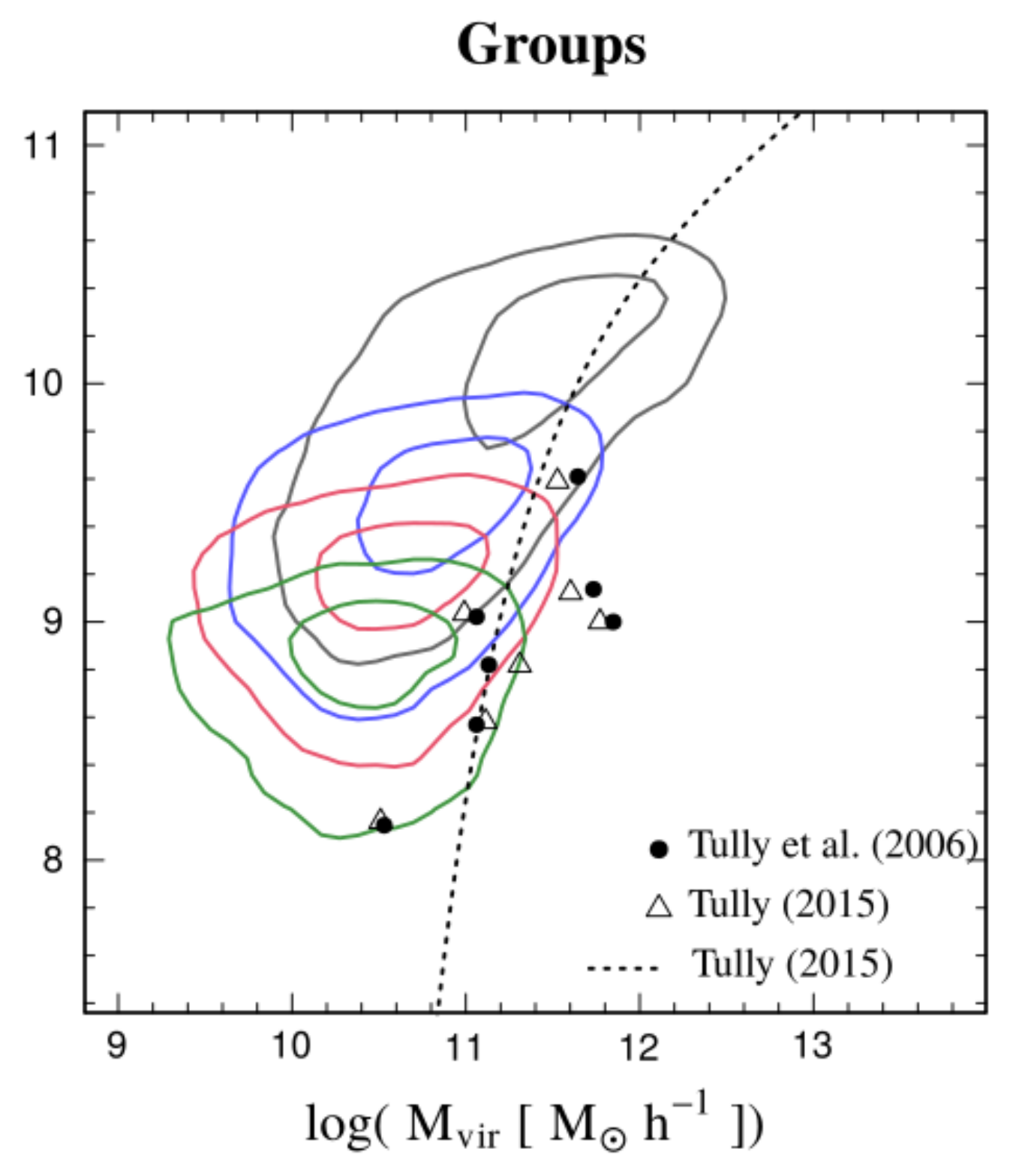}}
    \end{subfigure}
    \caption{ {\em B}-band luminosity as a function of the virial mass for our three 
    samples of dwarf galaxy systems (green, red, blue lines) and the 
    complete sample (grey lines), which are splitted in three different sub--samples 
    defined according to the belonging of their galaxy members to main host haloes:
    {\em Associations in different--haloes} (left panels), {\em Associations in mixed--haloes} 
    (middle panels) and {\em Groups} (right panels).
    Lines and symbols have the same meaning as in Figure~\ref{fig:mvir_lum}.}
    \label{fig:mvir_lum_change}
\end{figure*}

We also analyse the variation of the relation $L_{\rm B}$ vs. $M_{\rm vir}$, when we select systems 
according to the belonging of their galaxy members to the same host halo or to different ones.
This is shown in Figure \ref{fig:mvir_lum_change} for the three--samples of 
dwarf galaxy systems (green, red, blue lines) and the complete sample (grey lines).
Lines, contour levels and symbols have the same meaning as in Figure \ref{fig:mvir_lum}.
As in Figure \ref{fig:change}, we split each sample in {\em Associations in different--haloes} 
(left panel), {\em Associations in mixed--haloes} (middle panel) and in 
{\em Groups} (right panel).
Even though the {\em B}--band luminosity of systems in the different sub--samples are quite 
similar between themselves and in good agreement with the values of observed associations of 
dwarf galaxies, it is not the case for their virial mass.
The dwarf galaxies systems in the {\em Groups} sub--sample present virial 
masses smaller than observed, since smaller size gives place to smaller estimation of 
their virial mass. 
Sample size not withstanding, Fig 7 and 8 are a strong demonstration that the systems 
of dwarf galaxies observed by \cite{Tully:2006} and under examination here, 
are likely to be associations of dwarfs and not groups of dwarfs that share a host halo.

\section{Conclusions} \label{S_conclusions}

We use the high--resolution dark matter only SMDPL simulation \citep{Klypin:2016} 
coupled to the SAG semi--analytical model \citep{Cora:2018} to study 
associations of dwarf galaxies in the \mbox{$\Lambda$CDM} cosmological context.
We identify galaxy systems using a FoF algorithm with a linking length 
equal to $0.4\,{\rm Mpc}\,h^{-1}$, chosen to reproduce the size of the observational
associations of dwarf galaxies presented by \cite{Tully:2006}.
We analyse three different samples using a maximum dwarf mass 
\mbox{$\log_{10}(M_{\rm max}[M_{\odot}\,h^{-1}]) = 8.5, 9.0$ and $9.5$} and 
find that the last one agrees slightly better with the observational results
than the other two.
Our systems of dwarf galaxies have between 3 and 42 galaxy members; 
as expected from the dark matter halo mass function, there are more systems 
with fewer members than with many members.
More than $75$ per cent of our dwarf galaxy systems have 3 or  4 members.

We compare our predictions with the observational dwarf 
galaxy associations presented by \cite{Tully:2006} and 
find reasonable agreement not only in their sizes but also in 
their velocity dispersion and virial masses.
In agreement with observational results, our systems have 
typical sizes of $\sim 0.2\,{\rm Mpc}\,h^{-1}$, 
velocity dispersion of $\sim 30 {\rm km\,s^{-1}} $ and virial 
mass of $\sim 10^{11} {\rm  M}_{\odot}\,h^{-1}$.
Taking advantage of the information provided by the numerical simulation, 
we identify if the galaxy members of each one of our systems are located in the same 
main host halo or if they are located in different dark matter haloes.
We find that more than 90 per cent of our dwarf galaxies systems have 
their galaxy members residing in different dark matter haloes. 
On the contrary, less than 10 per cent of our dwarf galaxies systems have 
their galaxy members residing in the same dark matter halo.
Moreover, in agreement with observational results, our associations show 
relative large mass--to--light ratio (100 < M/L < 1000) if we 
assume that are bound systems.
We are able to assert that our model of galaxy formation SAG applied to 
the high--resolution SMDPL simulation, consistent with a flat \mbox{$\Lambda {\rm CDM}$} 
cosmological model, is capable to reproduce the existence and properties 
of dwarf galaxies associations without any extra fine--tuning.
We compare our results with the only seven observational associations 
of dwarf galaxies reported so far, but this analysis is important in 
the light of coming surveys, such as Dark Energy Spectroscopic Instrument 
(DESI) and Legacy Survey of Space and Time (LSST). 
These surveys promise the most detailed map of the nearby universe,
from which we will be able to systematically study dwarf galaxies 
and the different systems that they inhabit.

\section*{Acknowledgements}

We are very grateful to the referee for the constructive and insightful 
report which helped to improve our paper substantially.
The SMDPL simulation has been performed at LRZ Munich within the project pr87yi. 
The CosmoSim database (www.cosmosim.org) used in this paper is a service by the 
Leibniz--Institute for Astrophysics Potsdam (AIP).
Our collaboration has been supported by the DFG grant GO 563/24-1.
This work has been partially supported by the Consejo de
Investigaciones Cient\'ificas y T\'ecnicas de la Rep\'ublica Argentina
(CONICET), the Secretar\'ia de Ciencia y T\'ecnica de la
Universidad Nacional de C\'ordoba (SeCyT), Agencia Nacional de 
Promoci\'on Cient\'ifica y Tecnol\'ogica (PICT 2016-1975), and Universidad Nacional de La Plata, Argentina.
NIL acknowledges financial support of the Project IDEXLYON at the University of Lyon under the Investments for the Future Program (ANR-16-IDEX-0005). NIL also acknowledge support from the joint Sino-German DFG research Project ``The Cosmic Web and its impact on galaxy formation and alignment'' (DFG-LI 2015/5-1).
CVM acknowledges financial support from the Max Planck Society through a Partner Group grant.
GY acknowledges financial support by  MINECO/FEDER under project grant AYA2015-63810-P and by MICIU/FEDER under project grant PGC2018-094975-C21

\section*{Data availability}

The data underlying this article will be shared on reasonable request to the corresponding author.

%%%%%%%%%%%%%%%%%%%% REFERENCES %%%%%%%%%%%%%%%%%%

% The best way to enter references is to use BibTeX:

\bibliographystyle{mnras}
\bibliography{Bibliography.bib}

%%%%%%%%%%%%%%%%%%%%%%%%%%%%%%%%%%%%%%%%%%%%%%%%%%

% Don't change these lines
\bsp	% typesetting comment
\label{lastpage}
\end{document}